\documentclass[aps,prl,notitlepage,reprint,superscriptaddress,nofootinbib,longbibliography]{revtex4-2}

\usepackage{graphicx}
\usepackage{dcolumn}
\usepackage{bm}
\usepackage{amsmath}
\usepackage{upgreek}
\usepackage[colorlinks,linkcolor=red,citecolor=blue,urlcolor=red]{hyperref}
\usepackage[utf8]{inputenc}
\usepackage[T1]{fontenc}
\usepackage{mathptmx}
\usepackage{gensymb}
\usepackage{ulem}
\usepackage[colorinlistoftodos]{todonotes}
\renewcommand{\t}[1]{\mathrm{#1}}
\begin{document}

	
	\title{Multimode radiation pressure actuation of a mechanical resonator using a spatial mode sorter}
    	
 	
	\author{Morgan E. Choi}
	\affiliation{Wyant College of Optical Sciences, University of Arizona, Tucson, AZ 85721, USA}

    \author{Daniel Allepuz-Requena}
	\affiliation{Wyant College of Optical Sciences, University of Arizona, Tucson, AZ 85721, USA}

    \author{Diego Torres-Barajas}
	\affiliation{Wyant College of Optical Sciences, University of Arizona, Tucson, AZ 85721, USA}

    \author{Neshat Sadafi}
	\affiliation{Wyant College of Optical Sciences, University of Arizona, Tucson, AZ 85721, USA}
	
	\author{Dalziel J. Wilson}
	\affiliation{Wyant College of Optical Sciences, University of Arizona, Tucson, AZ 85721, USA}
	
	\date{\today}
	\begin{abstract}
\textcolor{black}{Multimode optomechanical scattering is ubiquitous in free-space imaging protocols, but its radiation pressure counterpart is challenging to detect with extended objects. Here we demonstrate radiation pressure actuation of a nanomechanical membrane using dynamically structured light from a spatial mode demultiplexer (SPADE). Using single- and two-mode illumination, we directly image the position-dependent intermodal coupling between Hermite-Gauss (HG) modes as mediated by the membrane's motion. As an application, we use multimode radiation pressure feedback to simultaneously cool two nearly degenerate membrane modes. Our results are applicable to diverse mechanical systems and serve as a \mbox{stepping stone to SPADE-based quantum~optomechanics.}}
	\end{abstract}
	
	\maketitle

\color{black}
In free-space imaging of a vibrating object, information is encoded into the frequency and spatial mode of inelastically scattered photons \cite{tebbenjohannsOptimalPositionDetection2019,pluchar2025imaging}.  Momentum conservation requires that this interaction be accompanied by radiation pressure: if mechanical motion scatters light between spatial modes, then dynamically modulating the spatial mode of incident light can excite that motion. While the former effect is readily detected \cite{parisi2026optomechanical,cheng2026spatial,tavernarakis2025wavefront,tawfik2026optimizing}, observing and exploiting multimode radiation pressure has remained challenging for extended mechanical objects.

Spatial mode demultiplexing (SPADE) provides a natural platform for multimode optomechanical measurement and control. Widely explored for advanced imaging tasks such as exoplanet detection \cite{norris2020all, centenera2026first} and superresolution microscopy \cite{rouviere2024ultra}, recently SPADE has been applied to quantum-limited readout of tethered \cite{choi2025quantum, tavernarakis2025wavefront} and levitated \cite{dinter2026three} nanomechanical oscillators, exploiting advances in low-loss photonic-integrated and multi-plane light conversion (MPLC)-based SPADE technology \cite{labroille2014efficient,fontaine2019laguerre}. A key prospect is radiation-pressure quantum control of mechanical motion \cite{choi2025quantum,dinter2026three}, leveraging the reciprocal ability of SPADE to dynamically structure light at frequencies beyond the reach of conventional spatial light modulators. 

Here we demonstrate the radiation-pressure counterpart of multimode scattering on an extended object, using a spatial mode demultiplexer as a dynamic wavefront-shaping tool.  Specifically, we use a Hermite-Gauss (HG) mode MPLC  to illuminate a Si$_3$N$_4$ membrane with two-mode structured light spatially modulated near a mechanical resonance. By mapping the position-dependent radiation pressure response, we directly visualize intermodal coupling between HG modes mediated by the membrane's motion.  The coherence and spatial structure of this coupling enable multimode optomechanical control protocols. As an illustration, we imprint the measured displacements of two near-degenerate membrane modes onto the phases of two HG-mode pairs, enabling selective and simultaneous cooling of their thermal motion.

\begin{figure}[t!]
\includegraphics[width=0.76\columnwidth]{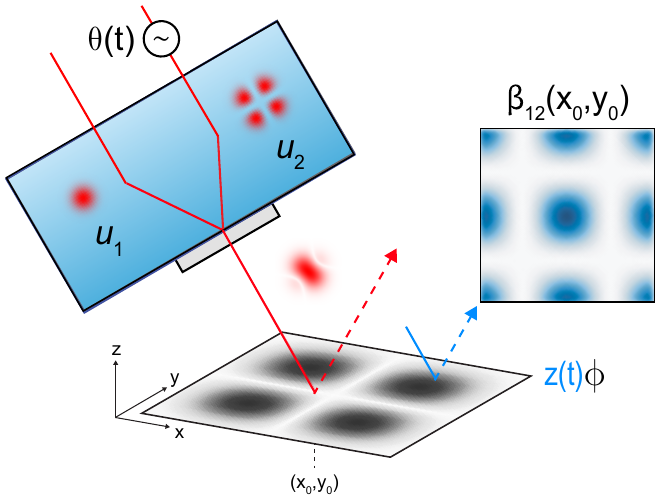}
\caption{\color{black}\textbf{Multimode radiation pressure actuation of a membrane.} A dynamically structured laser beam is produced by modulating the relative phase $\theta$ between two spatial modes $u_{1}$ and $u_{2}$ using a multi-plane light converter (MPLC). Reflecting the laser from a membrane (modeshape $\phi$) results in a vibration amplitude $z$ proportional to the optomechanical scattering coefficient $\beta_{12}=\langle u_2|\phi|u_1\rangle$.  Rastering the beam position $(x_0,y_0)$ yields a radiation pressure map $\beta_{12}(x_0,y_0)$.  } 
		\label{fig:1}
 \vspace{-1mm}
	\end{figure}
    
Figure \ref{fig:1} illustrates the principle of multimode radiation pressure actuation. A membrane with vibrational modeshape $\phi(x,y)$ is illuminated by  a laser with intensity $I(x,y,t) = |\sum_{mn}E_{mn}(t)u_{mn}(x,y)|^2$, dynamically structured by modulating the complex amplitudes $E_\t{mn}$ of individual HG components, with transverse modeshape $u_{mn}$. The generalized radiation-pressure force acting on the membrane mode is given by
\vspace{1mm}\begin{equation}\label{eq:1}
F_\t{RP}(t) \propto \langle I|\phi\rangle = \sum_{mn,kl}\beta_{mn,kl}E^*_{mn}(t)E_{kl}(t)
\end{equation}
where brackets $\langle\;\rangle$ indicate a transverse overlap integral and
\begin{equation}\label{eq:2}
\beta_{mn,kl} = \langle u_{mn}|\phi |u_{kl}\rangle
\end{equation}
are the optomechanical scattering coefficients.  The reciprocity of this interaction is readily seen for strong single-mode illumination ($u_\t{in}$): for a small membrane displacement~$z$, the reflected field is then, to first order,
\begin{equation}
E_\t{out} \simeq E_\t{in}u_\t{in}e^{2ikz\phi}
\simeq E_\t{in}\left(u_\t{in}+2ikz\sum_{mn}\beta_{mn,\t{in}}u_{mn}\right),
\end{equation}
where $k$ is the laser wavenumber and $\beta_{mn,\t{in}}=\langle u_{mn}|\phi|u_\t{in}\rangle$.

\begin{figure*}[t!]
    \centering
\includegraphics[width=2.05\columnwidth]{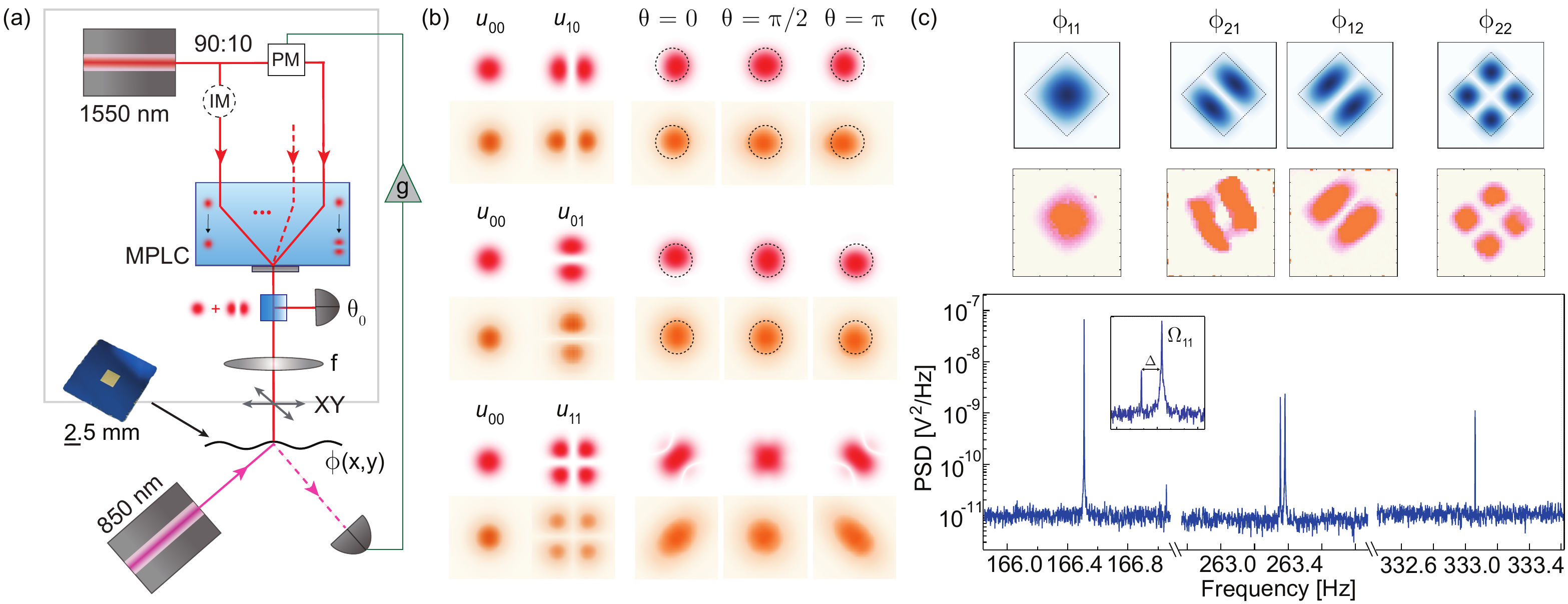}
    \caption{\textbf{Experimental setup and characterization of optical and mechanical modeshapes.} (a) Experimental setup for two mode radiation pressure actuation and feedback control. PM = phase modulator. MPLC = Multi-plane light converter. $f$ = lens for focusing onto membrane.  XY = micropositioning stage for rastering location of beam on membrane. Green line = feedback of photocurrent from optical lever (purple). Inset: photograph of 2.5 mm square Si$_3$N$_4$ membrane. (b) Intensity profiles of single-mode $u_{mn}$ and two-mode illumination fields $u_{00}+\alpha u_\t{mn}e^{i\theta}$ produced by the MPLC, for $|\alpha|^2\approx 0.1$ and $u_{mn}\in\{ u_{00},u_{21},u_{11}\}$.  Red = model, orange = measurement. 
    (c) Intensity profiles of membrane mechanical mode shapes $\phi_{ij}$ imaged by single-mode radiation pressure maps $\beta_{00,00}^2(x_0,y_0)\propto\phi^2(x_0,y_0)$ (Eq. 6).  Blue = model, orange = measurement.  The membrane is rotated by $\sim$45$\degree$ to match the orientation of the MPLC (Fig. 3).  Below: Optical lever photosignal power spectral density showing thermal displacement noise peaks near mechanical resonance (Eq. 5). The near-degenerate $\phi_{21,12}$ modes have a detuning of $\sim 30$ Hz. Inset: Zoom of the fundamental noise peak showing radition pressure tone detuned from resonance by $\Delta = 35$ Hz.}
    \label{fig:2}
\end{figure*}

We implement this scheme using a commercial HG-mode MPLC (Cailabs Proteus) and a high-Q Si\(_3\)N\(_4\) membrane resonator, adapting the apparatus described in \cite{choi2025quantum}.  As illustrated in Fig. 1, light from a 1550-nm laser is split by a 90:10 beamsplitter between HG$_{mn}$ and HG$_{kl}$ ports of the MPLC, operated as a fiber-to-free-space multiplexer.  The weak input is passed through a phase modulator $\theta(t)$, producing a purely spatially modulated (fixed total power) intensity pattern of the form
\begin{equation}\label{eq:4}
I(x,y,t) \propto u_{mn}^2+\alpha^2u_{kl}^2+2\alpha u_{mn}u_{kl}\cos[\theta(t)]
\end{equation}
where $\alpha^2 = |E_{kl}/E_{mn}|^2$ is the ratio of the HG$_{kl}$ an HG$_{mn}$ field intensities. Phase modulation $\theta(t) = \theta_0+\eta\cos(\Omega_0 t)$ near the resonance frequency $\Omega_\t{m}$ of a membrane mode then yields a displacement, here expressed as a power spectral density, of
\begin{equation}\label{eq:5}
    S_z(\Omega_0) = |\chi_\t{m}[\Omega_0]|^2 S_F^\t{RP}[\Omega] \propto \beta_{mn,kl}^2 \alpha^2\eta^2|\chi_\t{m}[\Omega_0]|^2,
\end{equation}
where $\chi_\t{m}$ is the mechanical susceptibility.

\color{black}

\color{black}
We first characterize two-mode illumination produced by the MPLC by imaging its transverse intensity profile $I(x,y)$ as a function of relative phase $\theta_0$. Figure~2 shows measurements taken with a beam profiler (Thorlabs BP209IR1) for HG$_{00}$ combined with HG$_{10}$, HG$_{01}$, and HG$_{11}$, together with the single mode profiles and the predictions of Eq.~(4).  Across these measurements, we observe fidelities $\langle I_\t{meas}|I\rangle\gtrsim 91\%$ using the mode waist and intensity ratio $\alpha^2$ as free parameters~\cite{SI}.

We next characterize the vibrational modeshape $\phi$ by mapping the amplitude response of the membrane to an intensity-modulated HG$_{00}$ beam.
The square Si$_3$N$_4$ membrane used in our experiment has a width of $L = 2.5\;\t{mm}$ and thickness of 90 nm, yielding drum modes $\phi_{ij}(x_0,y_0) = \sin(i\pi x_0/L)\sin(j\pi y_0/L)$ with measured resonance frequencies $\Omega^{ij}_\t{m}/(2\pi) \approx 166\,\t{kHz}\sqrt{(i^2+j^2)/2}$. Using a laser spot size $w_0$ smaller than the nodal spacing ($L/i$ and $L/j$) yields
\begin{equation}\label{eq:6}
S_z^{00,00}(\Omega_0\approx\Omega^{ij}_\t{m})\propto\langle u_{00}|\phi_{ij}|u_{00}\rangle^2\approx\phi_{ij}^2(x_0,y_0).
\end{equation}
where coordinates are chosen so that $(x_0,y_0)$ is the position of the laser beam on the membrane.  Rastering the laser beam across the membrane then gives a direct image of $\phi_{ij}^2(x_0,y_0)$.

Figure~\ref{fig:2}(c) shows HG$_{00}$ radiation-pressure maps (Eq.~\ref{eq:6}) that reproduce the expected squared mode shapes $\phi_{ij}^2(x_0,y_0)$.  The maps were acquired with the membrane housed in a high-vacuum chamber $(\sim10^{-8}\;\t{mbar})$ and a $w_0\approx 60\,\mu\t{m}$ beam rastered across its surface by translating the MPLC on a motorized  stage (Newport CMA-25CCCL). Displacement was read out with an auxiliary 850-nm optical lever~\cite{pluchar2025quantum} and membrane modes were identified by their $Q\sim 10^6$ thermal noise peaks [panel (b)]. For the measurements shown, we used a laser 
power $P_0\approx 3\,\t{mW}$ and modulation depth $\eta\sim1$. The modulation frequency was detuned by $\Omega_0-\Omega_\t{m}\sim 2\pi\times 10$ Hz to reduce sensitivity to photothermal resonance-frequency drift~\cite{tropper2026mode}, enabling thermal-noise-limited force measurements after several seconds of integration [panel (c), inset].

\begin{figure*}[ht!]
\includegraphics[width=1.7\columnwidth]{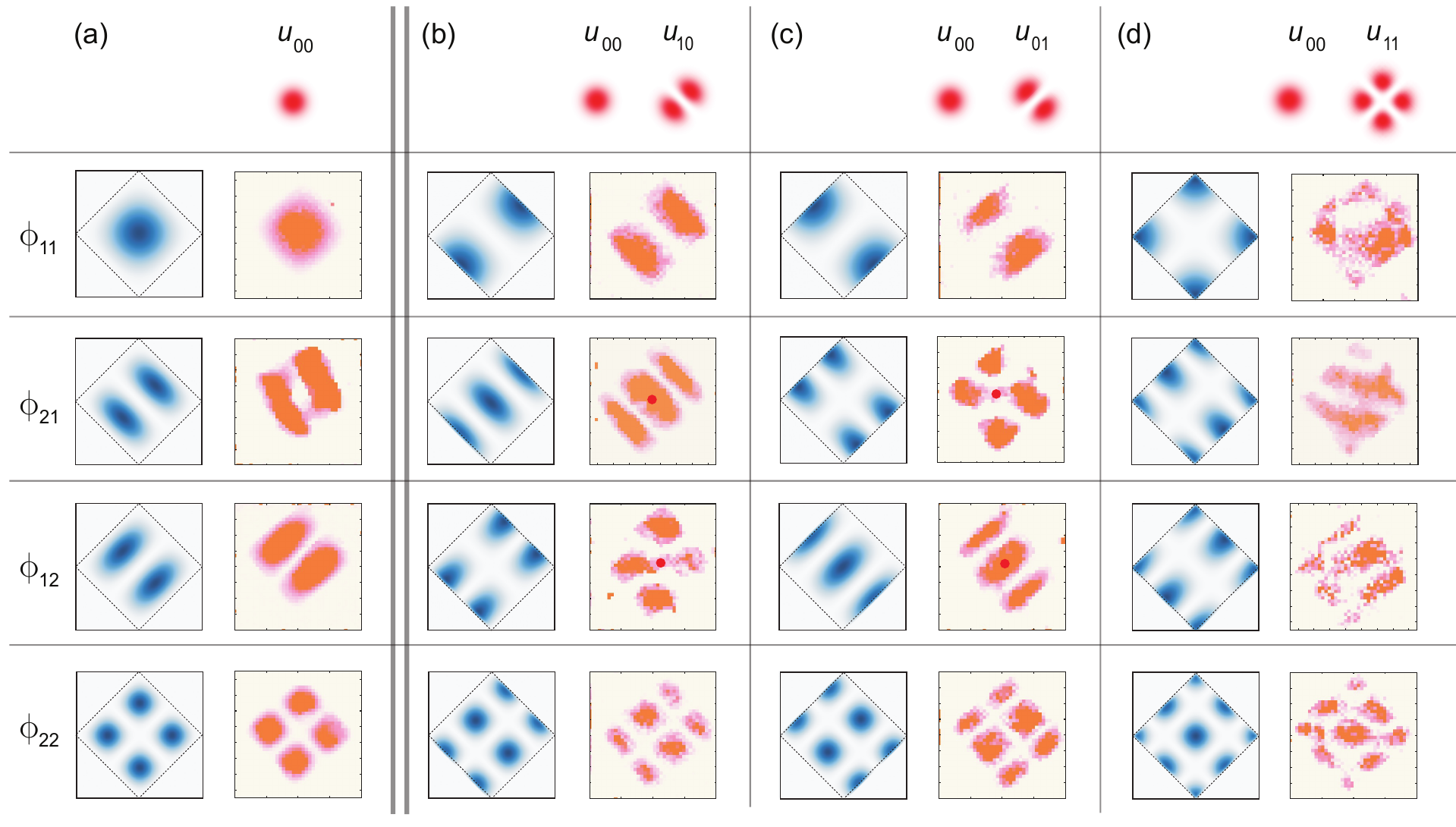}
		\caption{\textbf{Single and two-mode radiation pressure maps for HG-mode illumination of a square membrane.} Four mechanical modes $\phi_{ij}$ (rows) and illumination scenarios $u_{00}+\alpha u_{mn}e^{i\theta}$ (columns) are shown, corresponding to radiation pressure maps $\beta_{00,kl}^2(x,y)$.  Blue (orange) = model (measurement):
        (a) Single-mode $u_{00}$ maps $\beta_{00,00}^2(x,y)$ from Fig. \ref{fig:2}c. (b-d) Two-mode maps for $u_\t{mn} \in\{u_{10},u_{01},u_{11}\}$ using $\alpha^2\approx 0.1$ (b,c) and $0.5$ (d). Red dots in (b) and (c) show the beam position used for the feedback cooling data in Fig. \ref{fig:4}.}
        \label{fig:3}
	\end{figure*}
    
Having confirmed the illumination and mechanical modeshapes, we set out to observe multimode radiation pressure by driving the membrane with two-mode structured light (Eq. \ref{eq:4}).  Following Fig. \ref{fig:2}, we combined HG$_{00}$ with HG$_{10}$, HG$_{01}$, or HG$_{11}$, whose interference produces spatiotemporal radiation pressure patterns with distinct transverse symmetries: $x$-translation, $y$-translation in $xy$-shear, respectively. By rastering these patterns across the membrane and monitoring the displacement response (Eq. \ref{eq:5}), we measured their coupling to the $\phi_{11}$, $\phi_{12}$, $\phi_{21}$, and $\phi_{22}$ mechanical modes.

Figure \ref{fig:3} shows the resulting two-mode radiation-pressure maps, alongside the single-mode HG$_{00}$ maps from Fig. \ref{fig:2}c. \color{black} (For convenience, the HG mode coordinate system has been rotated 45$^\circ$ to match the orientation of the MPLC relative to the micropositioning stage.) 
\color{black} Measurements were obtained using the same total power and spot size as in Fig. \ref{fig:2}, a beamsplitting ratio $\alpha^2\approx  0.1$, and phase modulation depth $\eta = \frac{\pi V_\t{peak}}{V_\pi}\approx \pi/2$, averaging the spectrum over the offset phase $\theta_0$, which was allowed to drift by $\sim 1\;\t{rad} /\t{sec}$.  
For the weaker, higher-order mode images in Fig. \ref{fig:3}(d), beamsplitting ratio $\alpha \approx 0.5$ was used and traces had a dwell time of $\sim$30 seconds to mitigate resonance frequency drift effects.

A striking feature is the broken symmetry induced by two-mode interference. Whereas single-mode HG$_{00}$ actuation produces maps with the symmetry of the mechanical modeshape, $\beta_{00,00}^2\approx\phi_{ij}^2$, the intermodal radiation-pressure maps approximate spatial derivatives of the mechanical modeshape: $\beta_{00,10}^2\propto(\partial_x\phi_{ij})^2$, $\beta_{00,01}^2\propto(\partial_y\phi_{ij})^2$, and $\beta_{00,11}^2\propto(\partial_x\partial_y\phi_{ij})^2$.  More generally, to leading order in $w_0/L$,
\begin{equation}
\beta_{mn,kl} \propto \partial_x^{|m-k|}\partial_y^{|n-l|}\phi_{ij},
\end{equation}
providing access to higher spatial derivatives through higher-order HG-mode pairs, or equivalently, higher-order HG modes by scattering from higher-order mechanical modes. 

The coherent nature and nontrivial symmetry of intermodal coupling offer unique possibilities for optomechanical control. In particular, spatial-mode coupling provides a means of discriminating mechanical modes that are otherwise degenerate in frequency. A recent proposal exploits this principle for coherent or measurement-based multimode feedback control of a levitated particle, using a spatial mode demultiplexer simultaneously as detector and actuator \cite{dinter2026three}. More generally, intermodal optomechanical coupling offers a route to simultaneous ground-state cooling of multiple frequency-degenerate mechanical modes---an outstanding challenge in cavity optomechanics \cite{lai2020nonreciprocal,cao2025optomechanical}---by enabling independent readout and actuation of each mode with sufficient quantum efficiency.

\begin{figure}[t!]
    	\includegraphics[width=0.9\columnwidth]{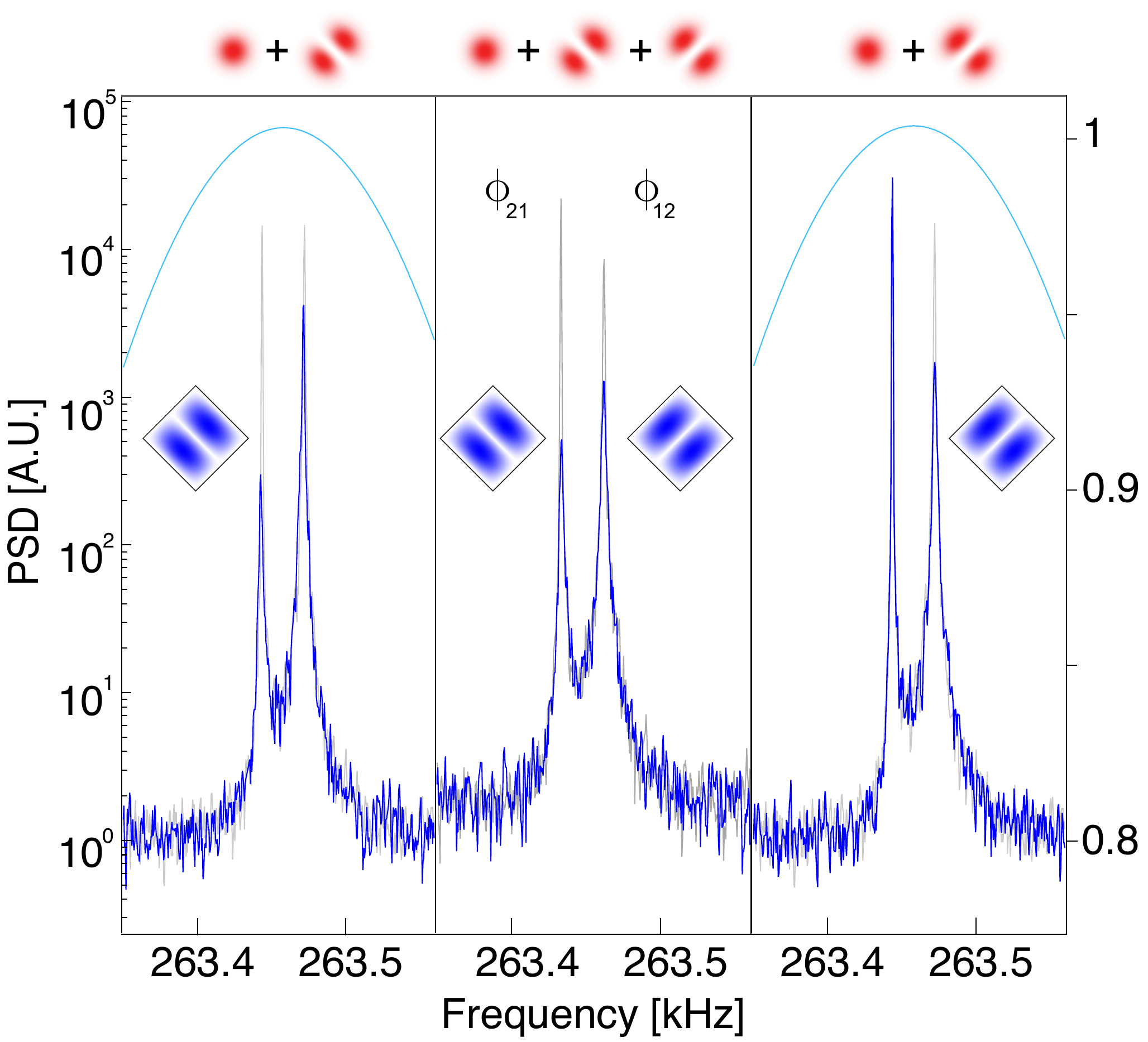}
		\caption{\textbf{Mechanical-mode-selective feedback cooling via multimode radiaton pressure actuation.}  Raw displacement power spectral densities (PSDs) for near-degenerature $\phi_{21}$ and $\phi_{21}$ membrane modes are shown, overlaid with the normalized magnitude of the feedback controller gain: a 1-kHz-wide notch filter (light blue, right axis). 
        Left (right) plot: selective cooling of the $\phi_{21}$ ($\phi_{12}$) mechanical mode via feedback to the relative phase of the $u_{10}$ ($u_{01}$) and $u_{00}$ optical modes. Middle plot: simultanous cooling of $\phi_{21}$ and $\phi_{12}$ via feedback to both $u_{10}$ and $u_{01}$. The beam position on the membrane is indicated by a red circle in~Fig.~3.}
		\label{fig:4}
	\end{figure}

As shown in Fig. \ref{fig:4}, we explore the actuator side of spatial-mode-resolved optomechanical control by simultaneously cooling the near-degenerate $\phi_{12}$ and $\phi_{21}$ membrane modes via HG$_{00/01}$ and HG$_{00/10}$ two-mode radiation-pressure~feedback
\begin{equation}\label{eq:8}
\begin{pmatrix}
F_\t{fb}^{(12)} \\
F_\t{fb}^{(21)}
\end{pmatrix}
\propto
\begin{pmatrix}
\beta^{(12)}_{00,10} & \beta^{(12)}_{00,01} \\
\beta^{(21)}_{00,10} & \beta^{(21)}_{00,01}
\end{pmatrix}
\begin{pmatrix}
\theta_{10} \\
\theta_{01}
\end{pmatrix}
\end{equation}
where here $F_\t{fb}^{(ij)}$ denotes the feedback force on mode $\phi_{ij}$, $\beta_{00,kl}^{(ij)} = \langle u_{00}|\phi_{ij}|u_{kl}\rangle$, and $\theta_{kl}$ is the feedback signal applied to the relative phase of the HG$_{00/kl}$ input field. To minimize crosstalk, following Fig. \ref{fig:3}, we align the input field near the membrane center, so that $\beta_{00,10}^{(12)}\approx\beta_{00,01}^{(21)}\gg \beta_{00,01}^{(12)},\;\beta_{00,10}^{(21)}$.  As expected from Eq. \ref{eq:8}, dissipative feedback of the optical lever signal $y$, $\theta_{kl}[\Omega_{ij}]\propto i\gamma_{fb}\Omega_{ij}y[\Omega_{ij}]$, selectively cold-damps the mode $\phi_{ij}$ for which $\beta_{00,kl}^{(ij)}\gg \beta_{00,kl}^{(ji)}$, while simultaneous cooling of $\phi_{12}$ and $\phi_{21}$ is achieved by simultaneous feedback to HG$_{00,01}$ and HG$_{00,10}$. In both cases, damping was limited to $\sim 10$ dB by electronic gain.   Details of the feedback circuit and offset phase-stabilization scheme are given in \cite{SI}.

In summary, we have used dynamically structured light from a spatial mode sorter to actuate the vibrational modes of a mechanical membrane resonator, directly visualizing the radiation-pressure counterpart of intermodal optomechanical scattering. We have also shown how multimode radiation-pressure feedback can be used to selectively and simultaneously cool nearly degenerate mechanical modes, addressing an outstanding challenge for multimode quantum state preparation \cite{lai2020nonreciprocal,cao2025optomechanical}. Looking forward, combining quantum-limited multimode readout \cite{choi2025quantum,dinter2026three} and radiation-pressure actuation of high-$Q$ levitated or suspended mechanical resonators \cite{engelsen2024ultrahigh} using SPADE technology could enable multimode ground-state cooling \cite{dinter2026three,lai2020nonreciprocal}, squeezed-state preparation \cite{marocco2026three}, and force microscopy \cite{schmerling2025optimal,hosseini2014multimode}. A key challenge is to scale such coherent optomechanical transceivers to larger numbers of spatial modes while maintaining high coupling efficiency and low crosstalk. High-speed reconfigurable mode sorters based on hybrid photonic-integrated circuits \cite{tang2018reconfigurable,boldin2026reconfigurable,butow2024generating} and piezoelectric spatial light modulators \cite{vanackere2025piezoelectrically} offer promising routes toward this goal, while adaptive wavefront-shaping strategies have already been explored for multimode cooling of three-dimensional particle arrays \cite{hupfl2023optimal}.  Multimode radiation-pressure actuation is also not restricted to HG modes: dynamic wavefront shaping in a Laguerre-Gaussian basis \cite{fontaine2019laguerre} could similarly generate rotating radiation-pressure profiles, extending recent optomechanical analysis of azimuthally structured light \cite{parisi2026optomechanical} to excitation and control of mechanical orbital-angular-momentum states~\cite{piotrowski2023simultaneous}.

\vspace{-2mm}
	\section*{Acknowledgements}
\vspace{-2mm}
 The authors thank Oscar Angulo for fabricating the mechanical membrane device, and Charlie Chisholm and Nathan Tankesly for engineering support. This work was supported by the National Science Foundation (NSF), award nos. 2239735 and 2606455, and by the Office of Naval Research MURI, award no. N000142612102. M.E.C. acknowledges support from the SPIE Women in Optics Scholarship. D.A-R., N.S., and D.T.B. acknowledge support from a Moore Foundation Postdoctoral Fellowship, the NSF ERC Center for Quantum Networks (CQN) grant number EEC-1941583, and the Friends of Tucson Optics scholarship, respectively.

\bibliography{ref}

\color{black}

	\end{document}


\linenumbers
	
	\title{
   Supplementary Information for\\ ``Multimode radiation pressure actuation of a mechanical resonator using a spatial mode sorter"}
 	
	\author{M. E. Choi}
	\affiliation{Wyant College of Optical Sciences, University of Arizona, Tucson, AZ 85721, USA}

    \author{D. Allepuz-Requena}
	\affiliation{Wyant College of Optical Sciences, University of Arizona, Tucson, AZ 85721, USA}

    \author{Diego Torres}
	\affiliation{Wyant College of Optical Sciences, University of Arizona, Tucson, AZ 85721, USA}

    \author{Neshat Sadafi}
	\affiliation{Wyant College of Optical Sciences, University of Arizona, Tucson, AZ 85721, USA}
	
	\author{D. J. Wilson}
	\affiliation{Wyant College of Optical Sciences, University of Arizona, Tucson, AZ 85721, USA}
	
	\date{\today}
	\begin{abstract}

	\end{abstract}
	
	\maketitle

\tableofcontents

\section{Image analysis and processing}
\begin{figure}[h]
    \centering
    \includegraphics[width=0.7\linewidth]{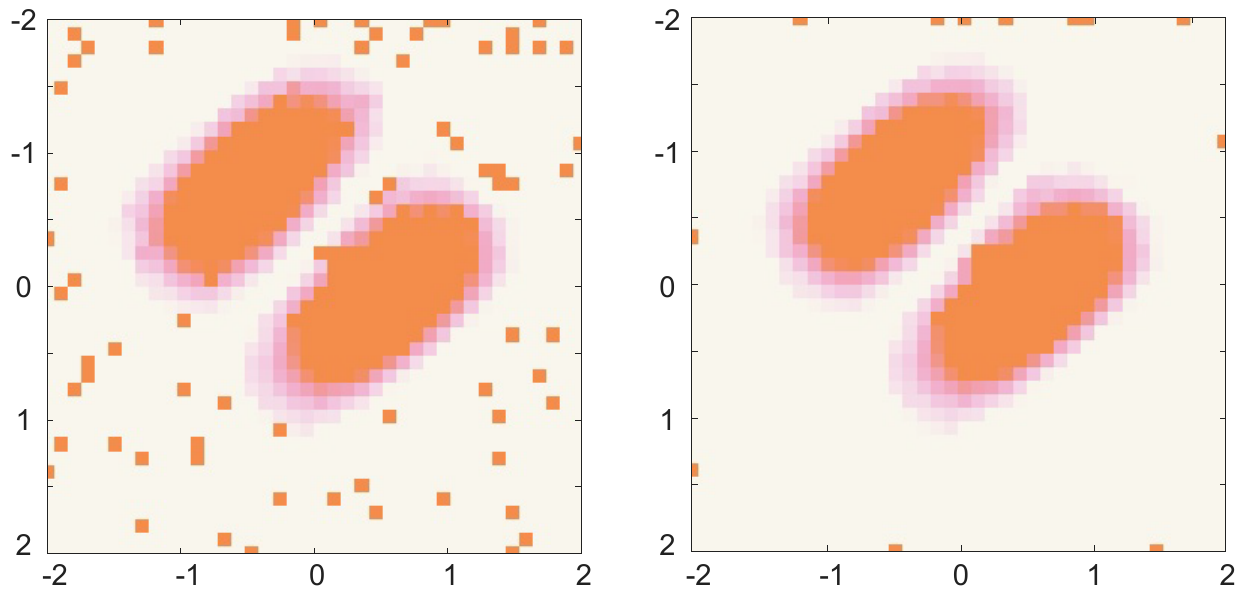}
    \caption{Example radiation pressure map before (left) and after (right) image processing. Ticks indicate millimeter steps.}
    \label{fig:cleaning}
\end{figure}

Due to thermal fluctuations in the mechanical resonance peak, our radiation pressure driven image backgrounds contain spurious peaks well above the background. These false peaks increases are artifact of our scan routine: We track the resonance of the mechanical peak, apply a detuned driving tone, and take the area of a fixed window around our expected driving frequency. Since the thermal resonance frequency shifts with temperature, and its amplitude fluctuates naturally, the window in which we calculate the area sometimes contains the thermal peak itself, causing an artificial hotspot on the final image. Similarly, sometimes within the real expected mode shape we see dips in the expected amplitude, from where the peak tracker lost the thermal peak and reassigned the window to an empty area, missing the drive~tone. 

To compensate for false peaks and smooth the background, we process the final dataset with a median filter. The filter compares each pixel in the image to its neighbors in a 3x3 window and, if the point is 1.25x larger or 0.75x smaller than its neighbors, replaces it with the median of the surrounding points. Since our mode shapes are characterized by gradual increases in amplitude rather than small local peaks, this method preserves the natural slopes of the lobes in our optomechanical coupling maps. Fig.~\ref{fig:cleaning} shows the raw and processed images for the $\phi_{12}$ single-mode illumination case. Another strategy to remove these peaks is to reduce their occurrence by a combination of longer averaging and adding a 30-60 second wait for thermalization in between steps, but these drastically increase the imaging time without improving the mode resolution. 


\section{Characterization of spatial modes}
\label{sec:spatial_mode_characterization}

Spatial modes generated by the multi-plane light converter (MPLC) were
characterized using a beam profiler (Thorlabs BP209-IR2/M) and compared with theoretical
Hermite--Gaussian (HG) distributions using spatial intensity correlation~\cite{rosales2017multiplexing}. The beam radii were obtained from
the measured fundamental mode, yielding
$w_x=965.1~\upmu\mathrm{m}$ and $w_y=984.9~\upmu\mathrm{m}$, and were fixed
for all subsequent mode reconstruction.

The HG field of order $(m,n)$ was modeled as
\begin{equation}
u_{mn}(x,y)
=
A_{mn}
H_m\left(\frac{\sqrt{2}x}{w_x}\right)
H_n\left(\frac{\sqrt{2}y}{w_y}\right)
e^{-\frac{x^2}{w_x^2}-\frac{y^2}{w_y^2}},
\label{eq:hg_mode}
\end{equation}
where $H_m$ and $H_n$ are Hermite polynomials and $A_{mn}$ is a normalization
constant. For each measurement, the experimental beam center was used as the
origin of the corresponding theoretical mode.

The agreement between the measured and theoretical intensity distributions---referred to as the fidelity in the main text---was
quantified using the Pearson spatial intensity correlation
\begin{equation}
C \equiv
\frac{
\sum_{i,j}
(A_{ij}-\overline{A})(B_{ij}-\overline{B})
}{
\sqrt{
\sum_{i,j}(A_{ij}-\overline{A})^2
\sum_{i,j}(B_{ij}-\overline{B})^2
}
},
\label{eq:intensity_correlation}
\end{equation}
where $A_{ij}$ and $B_{ij}$ are the normalized experimental and modeled
intensities, respectively. This metric was applied to the individually
generated $u_{00}$, $u_{01}$, $u_{10}$, and $u_{11}$ modes.

Coherent superpositions of the fundamental mode with each higher-order mode
were subsequently characterized for relative phases
$\theta=0$, $\tfrac{\pi}{2}$, and $\pi$. Their fields were modeled as
\begin{equation}
E(x,y;\theta)
=
a u_{00}(x,y)
+
b e^{i(\theta+\phi_0)}u_{mn}(x,y),
\label{eq:two_mode_field}
\end{equation}
where $r=b/a$ is the relative field-amplitude ratio and $\phi_0$ accounts for
a common phase offset. With the beam radii and spatial calibration fixed,
$r$ and $\phi_0$ were jointly determined from the three phase measurements by
minimizing
\begin{equation}
(r^{*},\phi_0^{*})
=
\underset{r,\phi_0}{\operatorname{arg\,min}}
\sum_{\theta\in\{0,\frac{\pi}{2},\pi\}}
\frac{1}{N}
\sum_{i,j}
\left[
A_{ij}(\theta)
-
B_{ij}(\theta;r,\phi_0)
\right]^2.
\label{eq:two_mode_optimization}
\end{equation}


\begin{figure*}[!t]
    \centering
\includegraphics[width=0.95\textwidth]{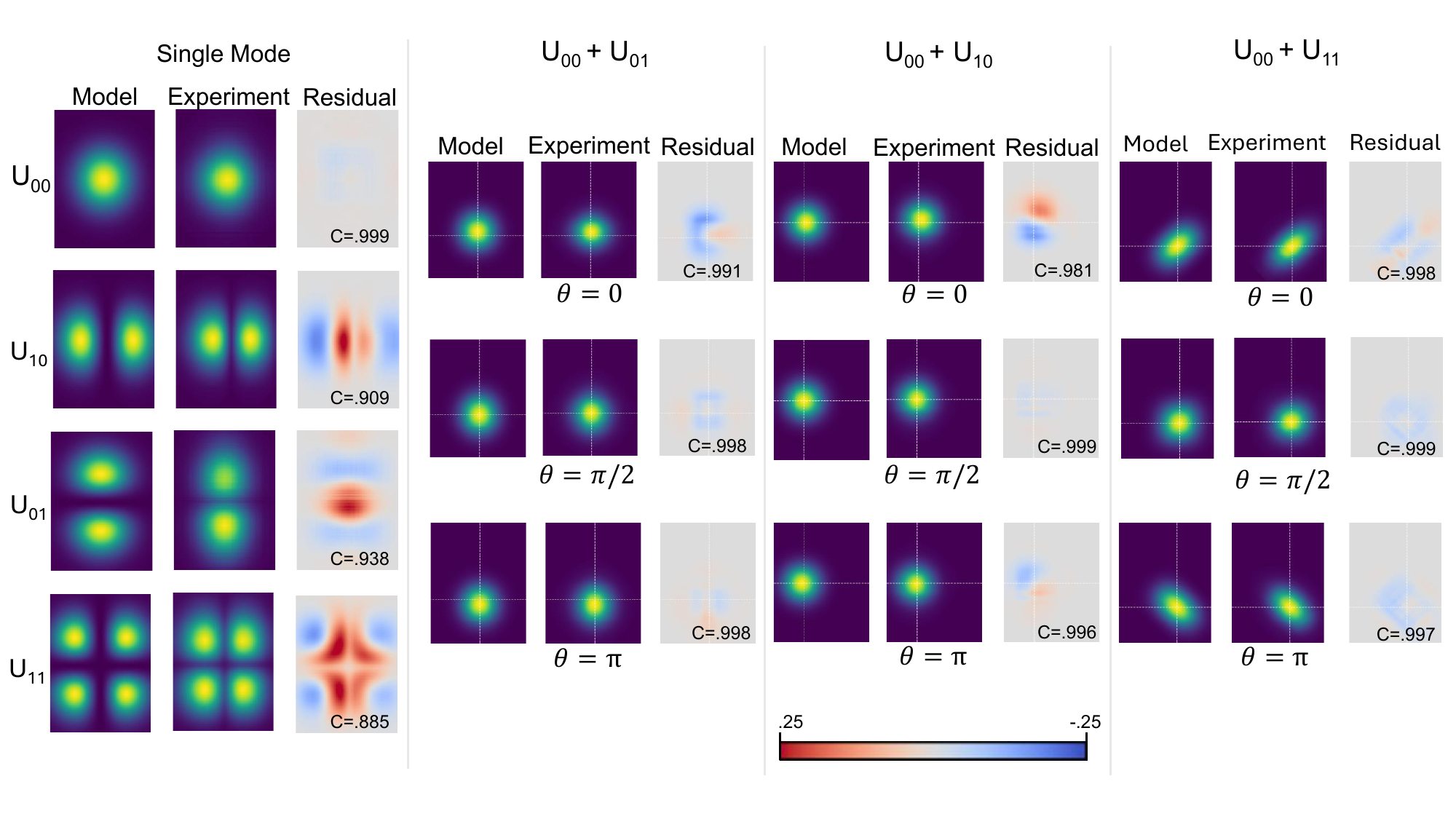}
    \caption{
    Characterization of MPLC-generated spatial modes. Single- (left) and two-mode superpositions (right) are compared with their theoretical models for relative phases $\theta=0$, $\tfrac{\pi}{2}$, and $\pi$. The spatial intensity correlation coefficient $C$ (Eq. \ref{eq:intensity_correlation}) quantifies their agreement.
    }
    \label{fig:HG_mode_fidelity}
\end{figure*}

Figure~\ref{fig:HG_mode_fidelity} summarizes the measured and modeled
individual single- and two-mode superpositions. The residual maps and
correlation coefficient $C$ quantify the agreement between experiment and
model. For the two-mode superpositions, the higher-order mode has a
substantially smaller field-amplitude weighting than the fundamental mode
($b\ll a$); consequently, the intensity distribution and its correlation
coefficient are dominated by the $u_{00}$ contribution.

\section{Feedback cooling via two-mode radiation pressure actuation}
\begin{figure}[b!]
    \includegraphics[width=0.75\columnwidth]{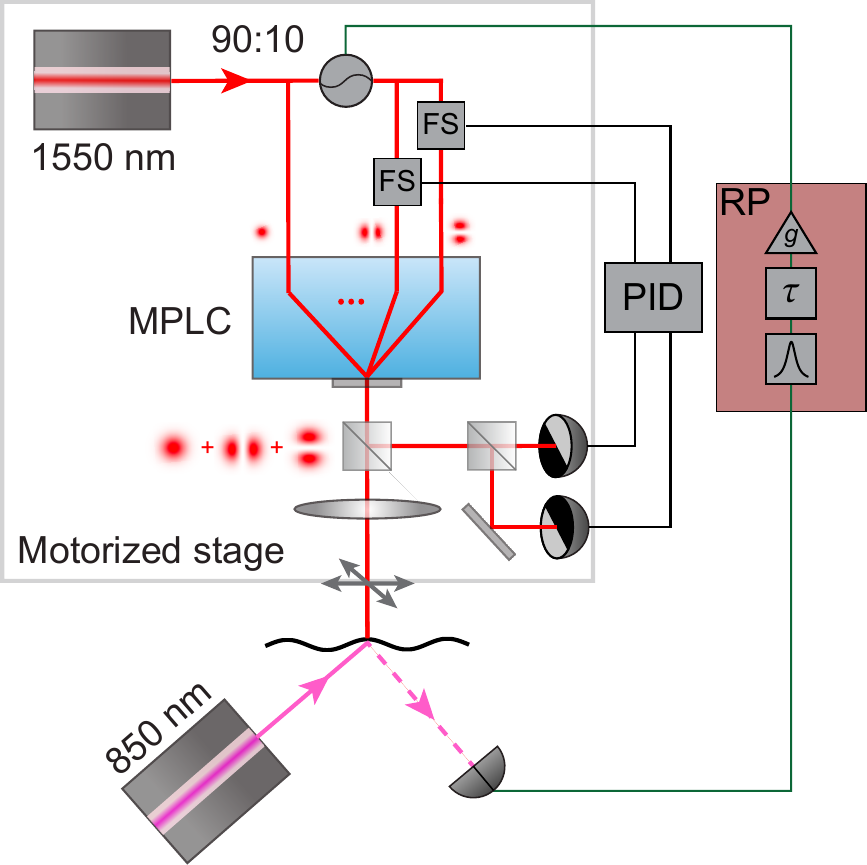}
    \caption{Measurement-based feedback cooling using a spatial mode sorter. The relative phase of the HG$_{00}$/HG$_{01}$ and HG$_{00}$/HG$_{10}$ fields is stabilized via a pair of fiber stretchers (FS). The control signal is obtained by tapping a fraction of the beam and measuring in with two partially blocked photodiodes. A damping force is engineered from the optical lever readout by means of an active bandpass filter and a delay implemented digitally on a Red Pitaya (RP). The resulting signal is fed into the high bandwidth fiber phase modulator.}
    \label{fig:cooling_setup}
\end{figure}
Our measurement-based feedback cooling scheme is based on the fact that, as demonstrated in the main text, modulating the relative phase $\theta$ between two distinct spatial modes $u_{mn}$ and $u_{kl}$ can generate a dynamical radiation pressure force:
\begin{equation}
    F_\mathrm{RP} (t) \propto \beta_{mn,kl}\cos{\left[ \theta_0 + \theta(t)\right]}.
    \label{eq:SI_radiation_pressure}
\end{equation}
where $\beta_{mn,kl} = \langle u_{mn}|\phi|u_\t{kl}\rangle$ is the spatial overlap with the mechanical mode $\phi$.
In order to use this force as a reliable actuator, we must stabilize the offset phase $\theta_0$ so that small changes in $\theta(t)$ transduce to linear changes in the feedback force. In our experiment, we simultaneously stabilize the relative phase between HG$_{00}$/HG$_{10}$ and HG$_{00}$/ HG$_{01}$. As shown in Fig. \ref{fig:cooling_setup}, this stabilization is achieved by measuring the interference between each mode pair with a partially blocked photodetector (a knife-edge approximation to a  split photodetector). This locking scheme will work as long as the overlap of the spatial modes is non-zero across the unobstructed area of the detector, viz., $\left\langle u_{mn} \vert f \vert u_{kl}\right\rangle \neq 0$ where $f(x,y)$ is the spatial sensitivity of the detector.

We can engineer the dynamics of radiation pressure by tailoring $\theta(t)$. In particular, cold damping occurs when the force opposes mechanical motion: $F_\mathrm{RP} \propto -\dot{z}$. The derivative $-\dot{z}$ can be estimated from the optical lever signal using a filter with transfer function $i \gamma_{fb} \Omega $. For a high quality resonator, a good approximation is a narrow bandpass filter with phase $\pi/4$ at its central frequency. Both filters effectively delay the signal by a quarter of a period which, for a sufficiently coherent resonator, is proportional to its current negative momentum. In our experiment, we apply a bandpass filter with adjustable gain and phase, resulting in a phase modulation:
\begin{equation}
    \theta(t) = -g y\left(t-\frac{\pi}{2\Omega_0}\right) \approx -g\dot{z}(t).
\end{equation}
According to Eq.~\label{eq:SI_radiation_pressure}, the radiation pressure force is:
\begin{equation}
    F_\mathrm{RP}(t) \propto -g\cos{\left( \theta_0\right)} \dot{z}(t).
\end{equation}
Because the phase $\theta_0$ determines both the sign and strength of the force, it is crucial to stabilize it. If the phase drifts, the feedback drifts from damping to anti-damping.\par

The feedback signal is generated digitally on a Red Pitaya FPGA board running the \textit{PyRPL} software~\cite{PyRPL}. A low-noise pre-amplifier (Stanford Research SR560) high-passes and amplifies the optical lever signal in order to fill the dynamical range of the Red Pitaya ADC. Once digitized, IQ demodulation and modulation is used to apply a Lorentzian bandpass filter, which is characterized using the internal network analyzer, and its amplitude response is shown in Fig.~4 of the main text. The filter is centered between the frequencies $\Omega_{12}$ and $\Omega_{21}$ and has an approximate 3~dB bandwith of 1~kHz. The phase of the filter is tuned so as to maximize cooling.
\newpage
\bibliography{ref}